\documentclass{elsart}

\usepackage{graphicx}
\usepackage{lineno}
\usepackage{epsfig}

\usepackage{amssymb}

\newcommand{\ind}[1]{_{\mathrm{#1}}}

\begin{document}

\begin{frontmatter}



\title{Venus wind map at cloud top level with the MTR/THEMIS visible spectrometer. \\I. Instrumental performance and first results. }


\author[cor1]{Patrick Gaulme}
\ead{Patrick.Gaulme@obspm.fr}
\author[cor2]{Fran\c{c}ois-Xavier Schmider}
\author[cor2]{Catherine Grec}
\author[cor3]{Arturo L\'opez Ariste}
\author[cor1]{Thomas Widemann}
\author[cor3]{Bernard Gelly}
\address[cor1]{LESIA, Observatoire
de Paris, 5 place J. Janssen, F-92195 Meudon cedex}
\address[cor2]{Laboratoire Fizeau, Universit\'e de Nice Sophia-Antipolis, CNRS-Observatoire de la C\^ote d'Azur, F-06108 Nice cedex 2}
\address[cor3]{THEMIS Observatory, La Laguna, Tenerife, Spain}

\begin{abstract}
Solar light gets scattered at cloud top level in Venus'
atmosphere, in the visible range, which corresponds to the
altitude of 67 km. We present Doppler velocity measurements
performed with the high resolution spectrometer MTR of the Solar
telescope THEMIS (Teide Observatory, Canary Island) on the sodium
D2 solar line (5890 \AA). Observations lasted only 49 min because
of cloudy weather. However, we could assess the instrumental velocity
sensitivity, 31 m s$^{-1}$ per pixel of 1 arcsec, and give a value of the amplitude of
zonal wind at equator at 151 $\pm 16$ m s$^{-1}$. 
\end{abstract}

\begin{keyword}
Venus\sep Wind\sep Clouds\sep Visible\sep  Spectrometry

\PACS
\end{keyword}
\end{frontmatter}
\linenumbers
\section{Introduction}
\label{intro}

ESA's Venus Express (VEx) space probe has been orbiting around
Venus since April 2005. The mission's main goal is a better
understanding of the atmospheric circulation, in particular of the
wind super-rotation. The key questions regard the meridian
circulation at top cloud level, the vertical extension of Hadley
cells and the latitudinal dependance of the zonal wind. VEx
obtains wind velocity map with cloud tracking and wind vertical
profile from thermal wind maps. VEx measurements are limited by
three factors. First, because of a very eccentrical orbit, i.e. a
strong velocity at periastron, VEx is not able to follow the
motion of cloud features, at latitude above $20^\circ$ N. Second,
with cloud tracking, the temporal resolution cannot get lower than
1 h. At last, wind measurements are restricted to two levels,
corresponding to cloud top, that is to say at 67 km on the day
side and 50 km on the night side (Drossart et al. 2007). That is
why, a large ground-based support has been organised around the
2007 Venus' maximum elongation in May-June and early November. The
main idea was to get radial velocity measurements with
spectrometry, almost simultaneously, in several spectral ranges in
order to probe as many levels as possible of the upper atmosphere.

Hereafter, we present spectrometric measurements on the sodium D2
solar line (5890 \AA), lead at THEMIS solar telescope (Teide
Observatory, Canary Islands), on discretional time, on November
7th 2007. The objective of this run was the evaluation of
instrumental performance of the MTR spectrometer, for radial
velocity measurements on a planetary target. The advantage of
MTR/THEMIS spectrometer with respect to point to point
echelle-spectrometer measurements is the ability to build velocity
maps, using a long slit entrance, greater than planetary diameter,
which allows us to scan completely the whole planet with only 15
positions of the slit. Because of cloudy weather, observations
lasted only 49 min. However, it is enough to evaluate the signal
to noise level and to give an estimate of the zonal wind velocity.
We present the instrument main properties (Sect. \ref{instru}),
the data processing (Sect. \ref{data_proc}) and, then, the
estimate of the instrumental performance and velocity field
(Sect. \ref{results}).

\section{Instrument main characteristics, observing conditions and expected performance} \label{instru}
\subsection{MTR-THEMIS high resolution spectrometer}
THEMIS (T\'elescope H\'eliographique pour l'Etude du Magn\'etisme
et des Instabilit\'es Solaires) is a French-Italian solar telescope
dedicated to accurate measurement of polarisation of solar
spectral lines, with high spatial, spectral and temporal
resolutions (Mein \& Rayrole 1985). It is a 90-cm diameter
Ritchey-Chr\'etien telescope. For present observations, it has
been operated in the MTR (MulTiRaies) mode (Rayrole \& Mein 1993)
which allows spectropolarimetric observations in up to 6 different
spectral domains simultaneously. For radial velocity measurements,
the polarimetric analysis was skipped and only spectrometric
information is considered. For this test run, we used the existing setup of the instrument and focused on the sodium D2 solar line
(5890 \AA). It is one of the deeper lines in the optimal
spectral domain for MTR detectors, but it is not optimum for Doppler sensitivity. We might search for a better line for future observations.

\subsection{Observing conditions}
Data were acquired on November 7th, 2007 between 13:06:52 h and
13:55:34 h (UT). The slit entrance dimension was about $(0.5 \times
100)$ arcsec. The spectral resolution was about 20 m\AA, while the
seeing has been estimated to 1 arcsec. The detector is a $512\times
512$ pixels CCD. The spatial dispersion upon the detector was
equal to 0.2 arcsec pixel$^{-1}$ and the spectral dispersion was
equal to 11.7 m\AA \ pixel$^{-1}$. The exposure time has been set to 10 s, while the readout time was less than 50 ms per exposure. We have considered the latter as negligible.

\begin{figure}
\hspace{2.5cm}
\includegraphics[width=8cm]{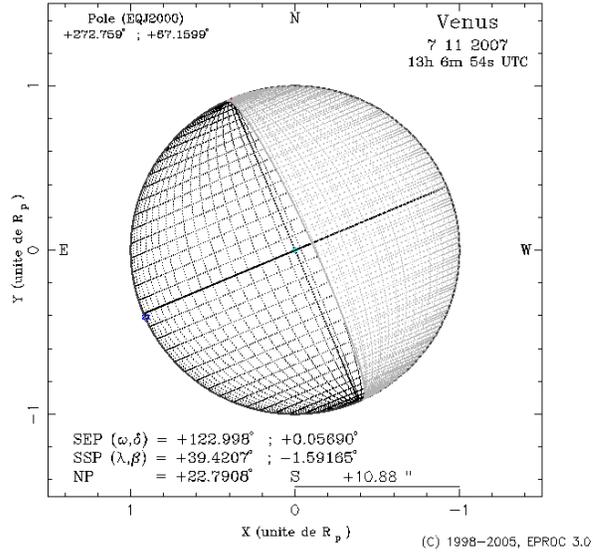}
\caption{Venus appearance during observations, on November 7th,
2007 at 13:06:54 h (UTC). The planetary radius is about 10.88
arcsec and the phase angle about $83.59^\circ$. SEP and SSP stand
for sub-Earth point and sub-solar point.} \label{fig:venus_imcce}
\end{figure}

The planet's diameter was about $21.76$ arcsec and the phase angle
about $83.59^\circ$ (Fig \ref{fig:venus_imcce}), which was close
to maximum elongation (on October 27th, 2007). Only 15 scans
regularly spaced of about 0.8 arcsec were necessary to map the whole enlightened part of the planet.
However, because of bad weather conditions, the scanning schedule
got changed and did not work nicely. In particular, no dark field
nor flat field were done and the scanning was off. The guiding was
manually set on Venus' limb, and the positioning upon the disk slowly
drifted along the run. Therefore, the planetary scan was only due
to the drift of the planet inside the field of view, which has
limited the coverage to 70 \% of the radius, i.e. an extension of about 8 arcsec along the equator (Fig. \ref{fig:scan}).
Nevertheless, 318 spectra were obtained in the 49-minute observation
run. Observing conditions were satisfactory during acquisition as
illustrated by the stability of the mean intensity of the
terrestrial signal (Fig. \ref{fig:mean_intensity}). We could
evaluate the instrumental performance and estimate qualitatively the
velocity range along the scan. The terminator region, where
the sun-Venus Doppler effect is expected to be maximum, is covered by
the observation. Expected performance and detailed analysis are presented in next sections.

\subsection{Theoretical performance}
\label{theo_perf} 
The principle of our observations rests on the measurement of the position of a solar Fraunhoffer line (D2, sodium),  which gets shifted by Doppler effect after reflection on Venus cloud decks. The radial velocity sensitivity depends on the sodium line thickness and the total amount of photons. First, measured on highly resolved spectrum, the local slope of the D2 sodium line appears to be $(\delta I/I)_{\mbox{max}} = 10^{-4}$ per m s$^{-1}$ at maximum, or, in average $<\delta I/I> = 0.5\ 10^{-4}$ per m s$^{-1}$ for two 60 m\AA\ bandwidths at each side of the line (see Fig. \ref{fig:sensibilite_D2}). Second, knowing that Venus emits almost 4.8 10$^{11}$ photons s$^{-1}$ m$^{-2}$ (1000 \AA)$^{-1}$ in the visible range on a 187-arcsec square lighted surface, that the telescope's efficient surface is about $0.5$ m$^2$, that the slit width is open at 0.5 arcsec, and that the global transmission is about 5 \% at 5500 \AA, we expect 1156 efficient photons s$^{-1}$ arcsec$^{-2}$ on Venus.  Since the total duration of the run is about 49 min and the spatial extension of the observed zone is about 8 arcsec, almost 6.1 min are dedicated per each 1-arcsec position of the slit, i.e. for each Venus slab. Consequently, the total amount of photons per 1 arcsec square is almost 4.23 10$^{5}$, that is to say a signal to noise ratio SNR $\simeq$ 650. Therefore, the expected 1-$\sigma$ velocity sensitivity is about $1/(650 \times 0.5\ 10^{-4}) = 30.7$ m s$^{-1}$ per pixels of 1 arcsec square. 

\section{Data analysis} \label{data_proc}
\subsection{Cleaning out raw spectra}\label{sect:clean_out}

Fig. \ref{fig:spectre_brut} presents a typical raw spectrum obtained
on Venus. Y-axis corresponds to spatial dimension,
parallel to the terminator, and $x$-axis
corresponds to spectral dimension . The spatial range corresponds
to 100 arcsec and the total spectral range to 6 \AA. Since the entrance
slit is much larger than the planetary diameter, most of the
image is occupied by solar radiation scattered by Earth's atmosphere.
The Doppler shift of the D2 sodium line between Venus and Earth
atmosphere is clearly visible at image center on Fig. 5 (left).
Thinner lines crossing vertically
 the detector are telluric
absorption lines. A slight distorsion is visible across the field. This is a
consequence of the optical design of the spectrometer. At first
order, distorsion parameters can be considered uniform across
the field. We calculate them by fitting the Earth's D2 sodium line
with a second order polynomial. The distorsion is then rectified
with a cubic spline interpolation algorithm. This
effect is purely instrumental. Distorsion can be considered constant
through the observation run, so the fit has to be performed
only once, on one image. The same correction algorithm is
is applied to all spectra.

\subsection{Positioning on the planetary disk}

\begin{figure}
\hspace{2cm}
\includegraphics[width=9cm]{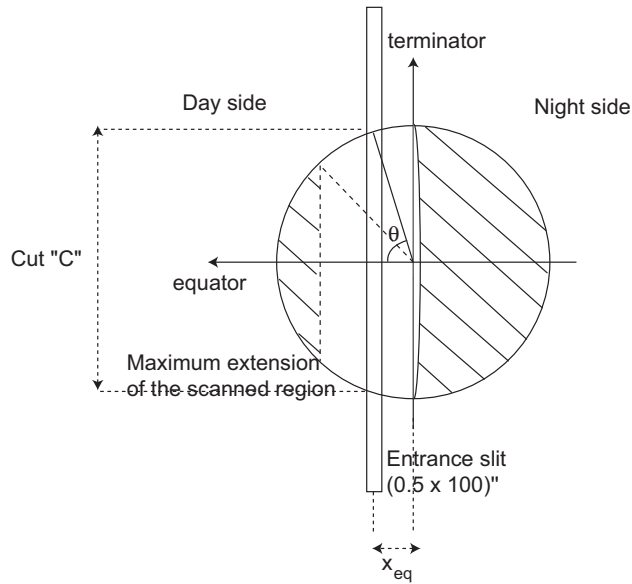}
\caption{Schematic review of notations used in this paper and
illustration of the spatial coverage of Venus by our observations.
The slanted lines indicate the region which is not covered by
observations. $\theta$ indicates the latitude and $C$ the cut
along the planet, through the entrance slit (in pixels). The
maximum extension of the spatial coverage reaches $\theta =
45^\circ$ on the planetary limb.} \label{fig:scan}
\end{figure}

\begin{figure}
\hspace{2.5cm}
\includegraphics[width=9cm]{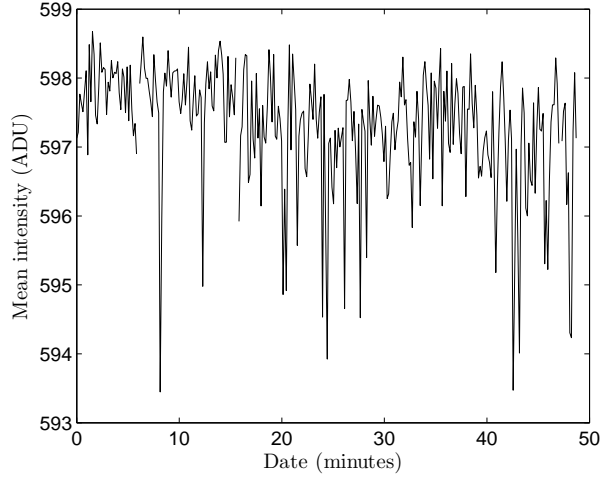}
\caption{Mean intensity measured on Earth's skylight background on
each spectrum of the temporal series. The mean value is equal to
597.6 ADU, whereas the standard deviation of the points is equal
to $\Sigma = 58.16$ ADU.} \label{fig:mean_intensity}
\end{figure}

\begin{figure}
\hspace{2.5cm}
\includegraphics[width=9cm]{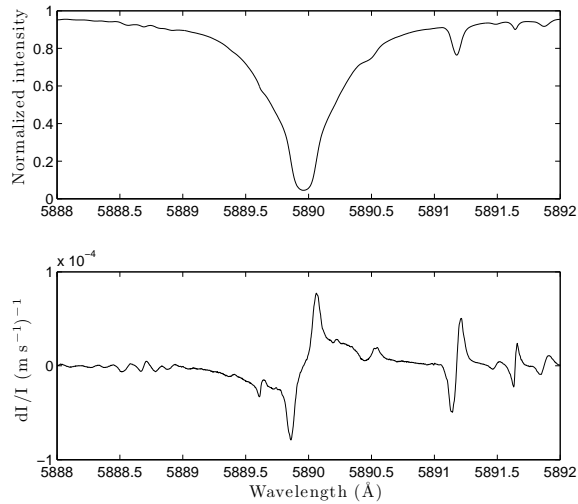}
\caption{Top: D2 sodium line on Sun spectrum as function of the wavelength (BASS 2000 database, http://bass2000.obspm.fr). Intensity has been normalized to 1. The Doppler sensitivity is related to the slope of the considered line. It reaches its maximum at the transmission level of 30\%. Bottom:  the slope of the upper figure, converted in meter per second. In average, the Doppler velocity sensitivity is about $<\delta I/I> = 0.5\ 10^{-4}$ per m s$^{-1}$ for a bandwidth of 60 m\AA.} \label{fig:sensibilite_D2}
\end{figure}

\begin{figure}
\includegraphics[width=7cm]{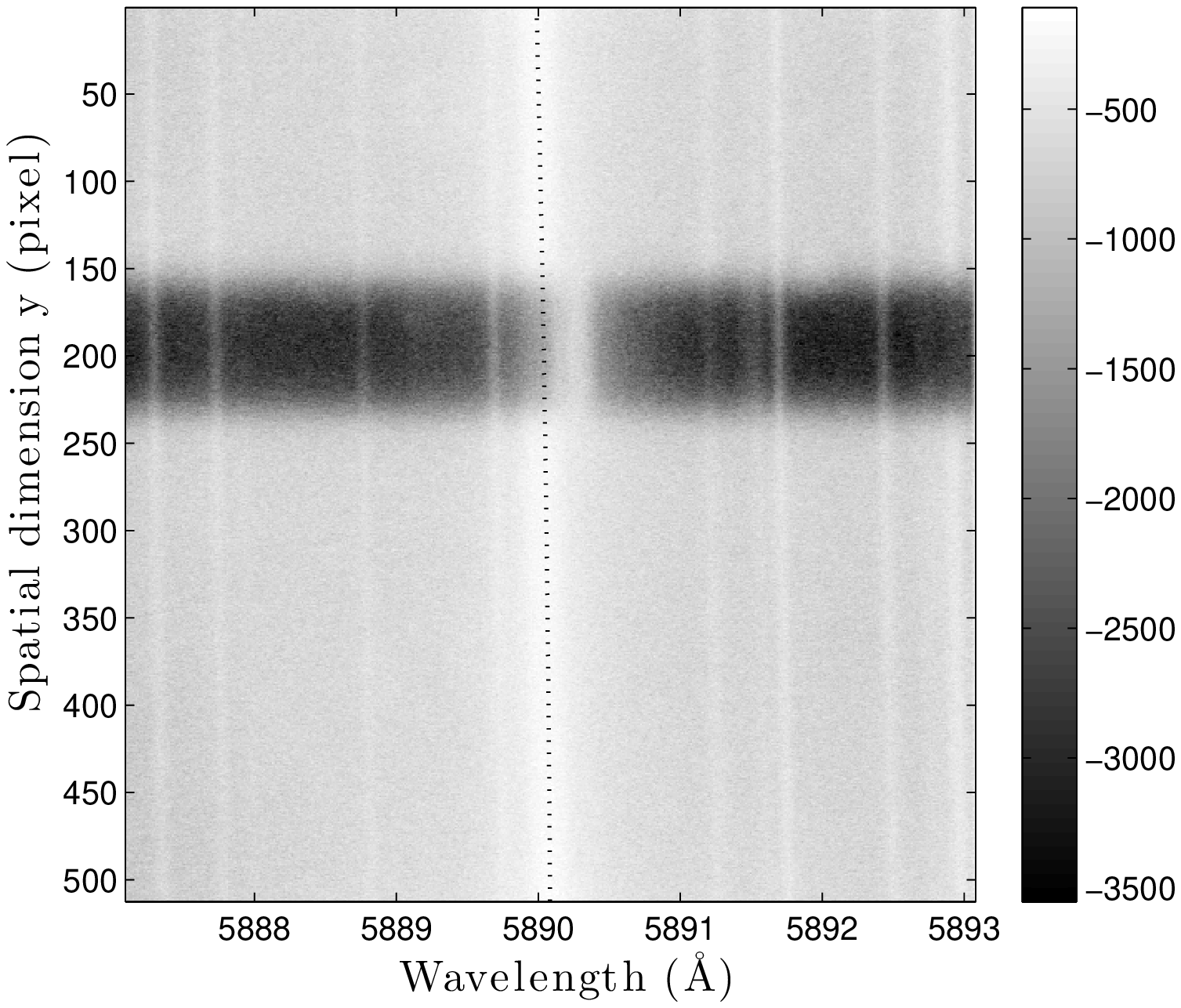}
\includegraphics[width=7cm]{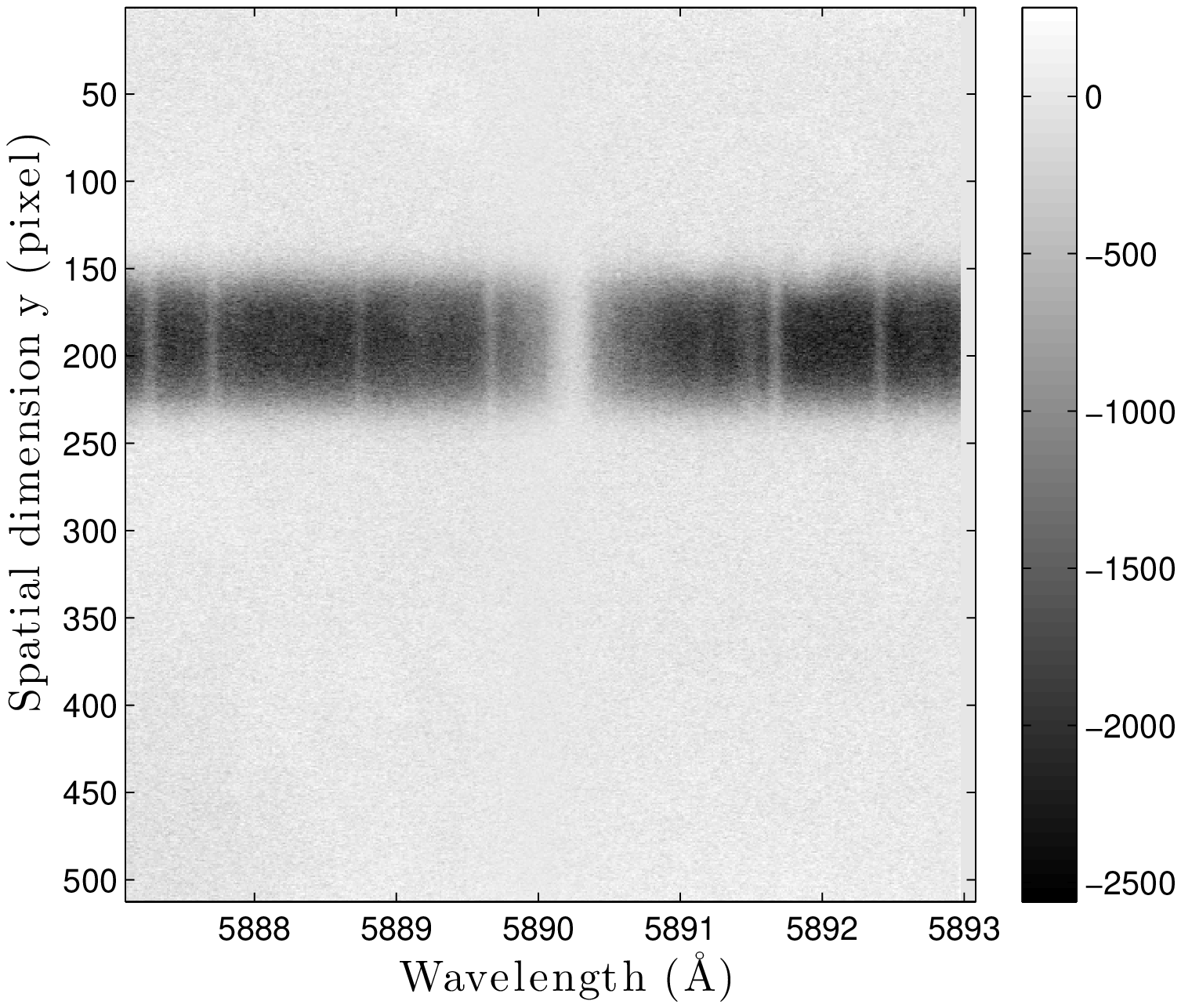}
\caption{Left: raw spectrum of Venus, centered on D2 sodium solar
line. The $y$-axis corresponds the spatial dimension, while the
$x$-axis squares with the spectral dimension. Venus corresponds to
the dark region on the detector, with highest intensity, whereas
the light background corresponds to the Earth's sky spectrum. The
D2 line on Venus is clearly shifted with respect to the Earth's.
The slight curvature across the whole image is estimated by
fitting the D2 sodium line scattered by Earth's atmosphere with a
second order polynomial (dot line). Right: clean spectrum. After
straightening out the distorsion, skylight sodium light has been
averaged over the background, and subtracted to Venus. Thinner
lines correspond to telluric absorption lines.}
\label{fig:spectre_brut}
\end{figure}

The main cause of uncertainty in processing the data comes
from the positioning on the disk. Indeed, as the
quick scan of the planet did not work, we have no direct
measurement of the absolute position. The positioning is
determined as a relative function of the initial pointing
along the terminator. The spatial scale is determined using the
spatial extension of Venus on first spectra, which covers
115 pixels, and fits the expected diameter of Venus upon
the detector ($21.76\times0.2=109$ pixel), taking into account the
seeing effect. If we suppose that the slit slewed parallel to the
terminator, the positioning upon the planet only depends on the
ratio of the measured cut along the planet with respect to its
size at terminator.
\begin{equation}
x\ind{eq} = \cos \theta \ \ \mbox{with}\ \ \theta = C/D\ind{venus}
\label{eq:position}
\end{equation}
where $x\ind{eq}$ indicates the $x$-coordinate along the planetary
equator, $\theta$ the latitude, $C$ the extension of the planetary
cut (pixels) and $D\ind{venus}$ Venus measured diameter (pixels)
(see Fig \ref{fig:scan}). Note that this expression assumes that
Venus was at quadrature (phase angle $90^\circ$) and the central meridian is
the terminator. The phase angle was $84^\circ$ instead of $90^\circ$,
yielding a difference of 1.1 arcsec, that is the spatial resolution by taking into account the seeing.

\begin{figure}
\hspace{0.7cm}
\includegraphics[width=12cm]{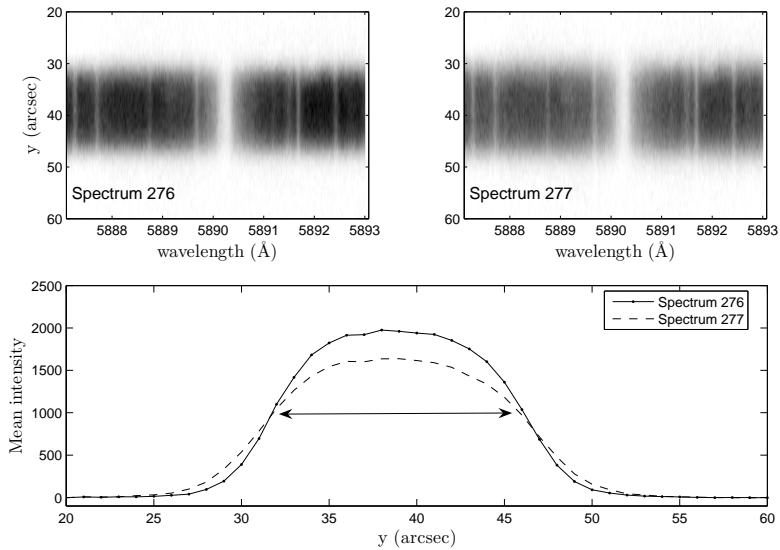}
\caption{Top: two consecutive spectra (labelled 276 and 277 among
the 318 spectra temporal series), which have been acquired with an
interval of 10 s. Bottom: projection of both spectra along the
$y$-axis, in order to evaluate the spatial extension of the planet
selected by the entrance slit. Width determination is the only
method to locate the slit projection on Venus.} \label{fig:seuil}
\end{figure}

The spatial extension of Venus on the detector is determined after
subtraction of Earth's skylight mean signal (Fig
\ref{fig:spectre_brut}). Then, the spectral image is projected
along the $y$-axis, in order to get a smooth spatial profile.
Venus' spatial dimension is arbitrarily defined as the region
where the intensity exceeds the 1/2 of the maximum value, i. e.
the full width at half maximum (FWHM) (see Fig \ref{fig:seuil}).
We estimate the spatial extension at half maximum in order to
minimize error due to seeing fluctuations, as illustrated by Fig.
\ref{fig:seuil}.

\begin{figure}
\hspace{3cm}
\includegraphics[width=8cm]{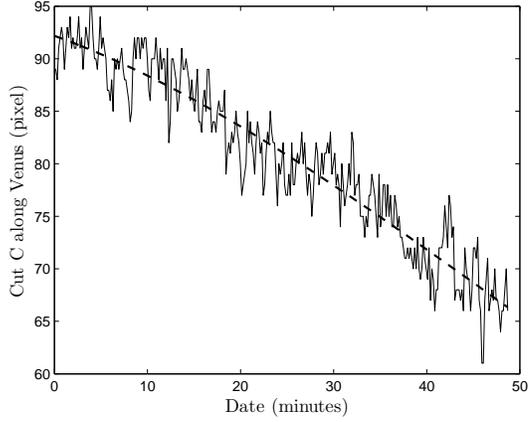}
\caption{Spatial extension estimate ``C'' as function of time,
along the observation run. The cut extension has been calculated
on full resolution images, in order to keep the original accuracy;
1 pixel corresponds to 0.2 arcsec. The extension is defined by the
width at half maximum. The solid line represents the measurement
of the spatial extension, while the dashed line represents the
polynomial fitting of Venus cut. The standard deviation of points
around the mean is equal to 2.46 pixels.} \label{fig:corde}
\end{figure}

Plotting the FWHM as a function of time shows a slow drift,
overlapped by a high frequency oscillation (Fig. \ref{fig:corde}).
Such fluctuations are due to the bias introduced by rapid seeing
($10-$s interval from one image to another). The fit value used to
calculate the position upon the planetary disk is obtained by a
3rd order polynomial. The standard deviation of the points with
respect to their smoothed profile is about 2.46 pixels, which
corresponds to 0.49 arcsec. This gives us the relative error bar
of the slit position estimate $x\ind{eq}$ on the disk (Fig.
\ref{fig:scan}).

\subsection{Doppler shift of D2 lines}

\begin{figure}
\hspace{2cm}
\includegraphics[width=10cm]{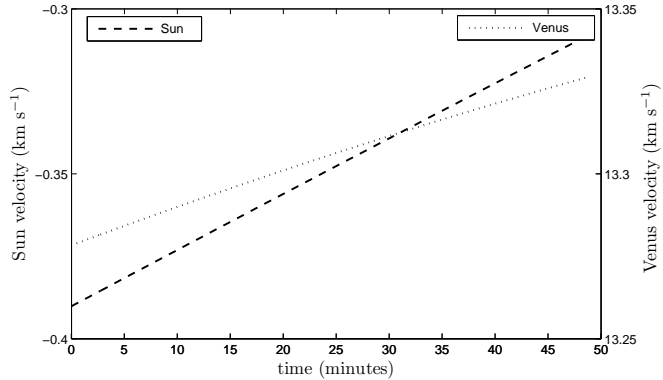}
\caption{Relative velocities with respect to Teide Observatory
between 13:06:52 h and 13:55:34 h (UT) on November, 7th 2007.
Initial date squares with 13:06:00 h, velocities are expresses in
km s$^{-1}$. Left $y$-axis indicates Sun relative velocity, which
mean amplitude is about $-0.35$ km s$^{-1}$. Right $y$-axis
indicates Venus relative velocity, which mean is about $+13.3$ km
s$^{-1}$. Consequently, the mean Doppler shift between D2 sodium
line scattered by Venus' and Earth' atmospheres is equal to
$+13.65$ km s$^{-1}$ (www.imcce.fr).} \label{fig:vitesses_derive}
\end{figure}

Doppler maps are obtained by measuring the shift between the D2
sodium lines, scattered by Earth's and Venus' atmospheres. The
shift must first be corrected from (i) Venus' center motion with
respect to observer and (ii) observer's motion with respect to the
Sun. All these components are well known, and get subtracted with
the help of ephemeris data (Fig. \ref{fig:vitesses_derive}).

\begin{figure}
\includegraphics[width=7cm]{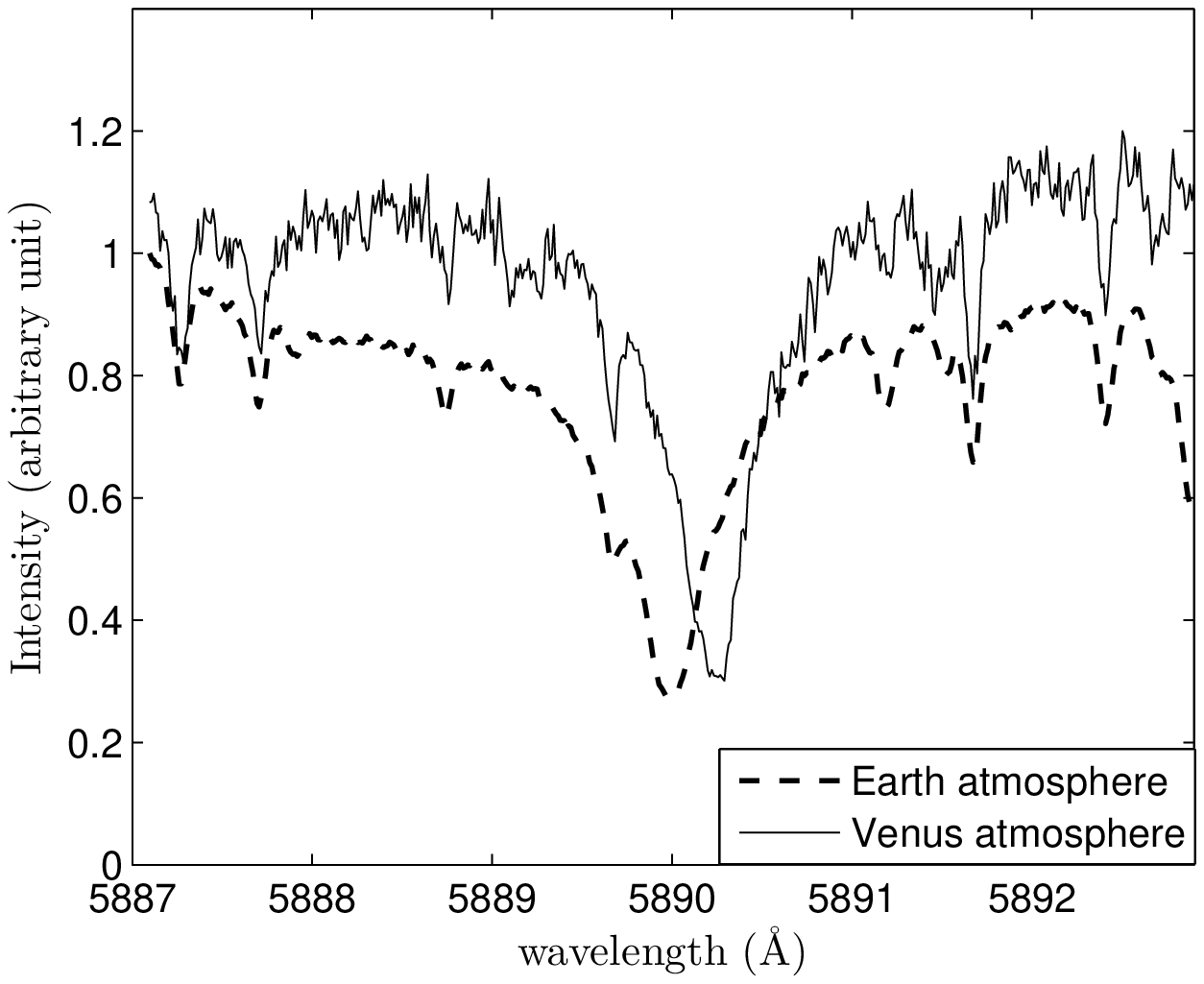}
\includegraphics[width=7cm]{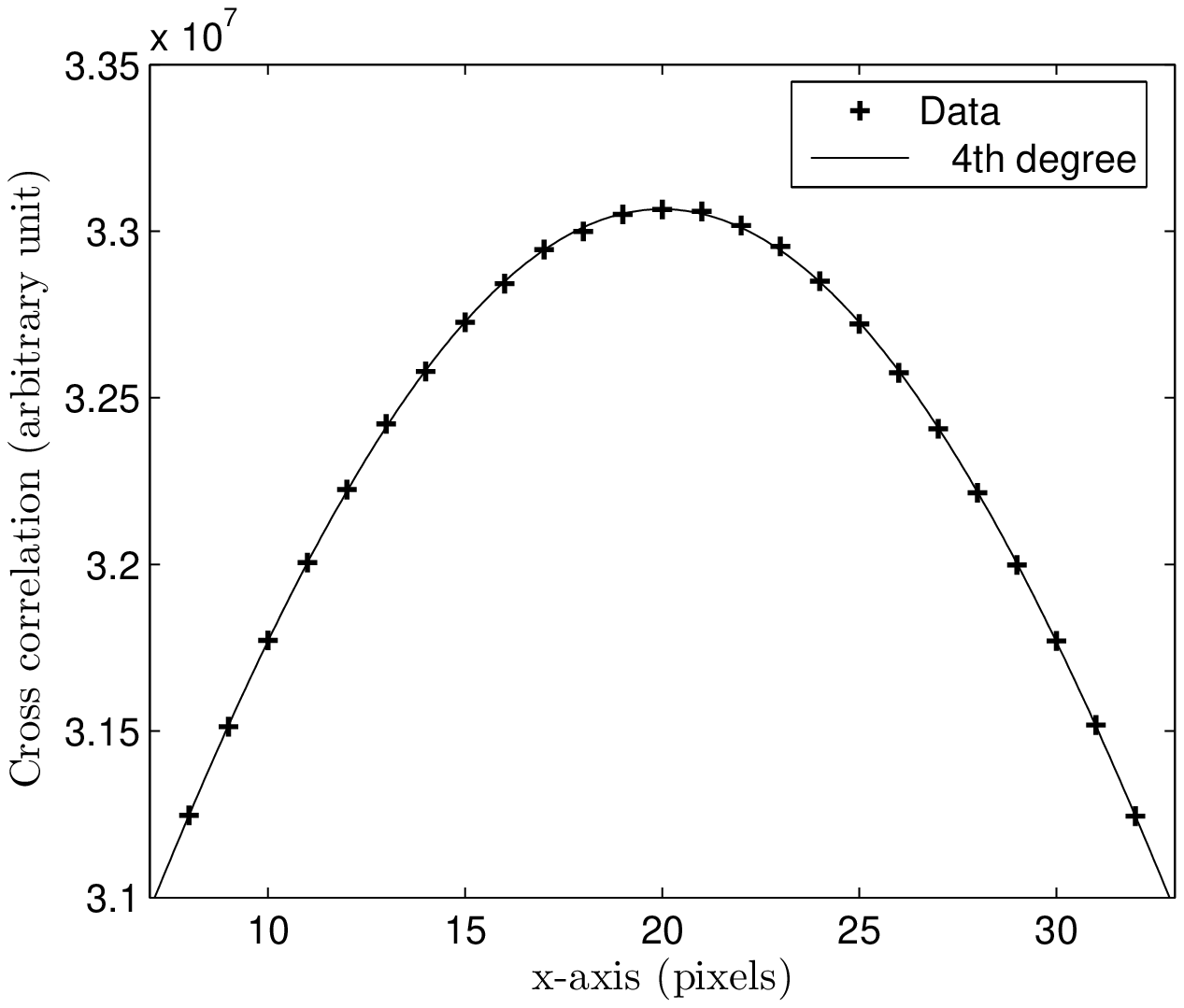}
\caption{Left: reference mean spectrum (dashed line) and Venus
spectrum as a function of the wavelength (full line). Reference
spectrum is calculated for each spectrum by averaging all the
skylight spectrum. Venus spectrum is obtained on a 1 arcsec
spatial resolution spectrum, and correspond to the planetary
equator. Both spectrum have been normalized with respect to their
maximum value and Venus spectrum has been offset, only for
graphical reasons. Right: 4th order polynomial fit of the maximum
of the cross correlation between Venus and reference spectra. The
Maximum position is equal to 20.51 pixels. }
\label{fig:cross_corr}
\end{figure}

The Earth scattered solar D2 line intensity is averaged over all
detector lines outside of Venus, to create a mean reference
spectrum. On Venus, the 0.2-arcsec lines are coadded by groups of
5 in order to reach a 1-arcsec vertical resolution along the slit,
thus improving the signal to noise ratio by a factor $\sqrt{5}$.
The reference spectrum is then correlated to the spectrum measured
on Venus, line by line; the Doppler shift corresponds to the
position of maximum value of the cross correlation. Points are
fitted with a 4th order polynomial, then, the maximum position is
determined by calculating numerically the zeros of the derivative
of the fitting function (Fig \ref{fig:cross_corr}). The fit
accuracy is strongly dependent upon S/N; that is why we consider
only the signal coming from Venus central region on the detector,
where amplitude is greater than half maximum amplitude, as for the
cut estimate (previous Sect.). In terms of angular size, it means
to keep 16 arcsec instead of 21.8 arcsec along the diameter; in
terms of latitude, it limits the map to $\pm 45^\circ$.

\begin{figure}
\center
\includegraphics[width=14cm]{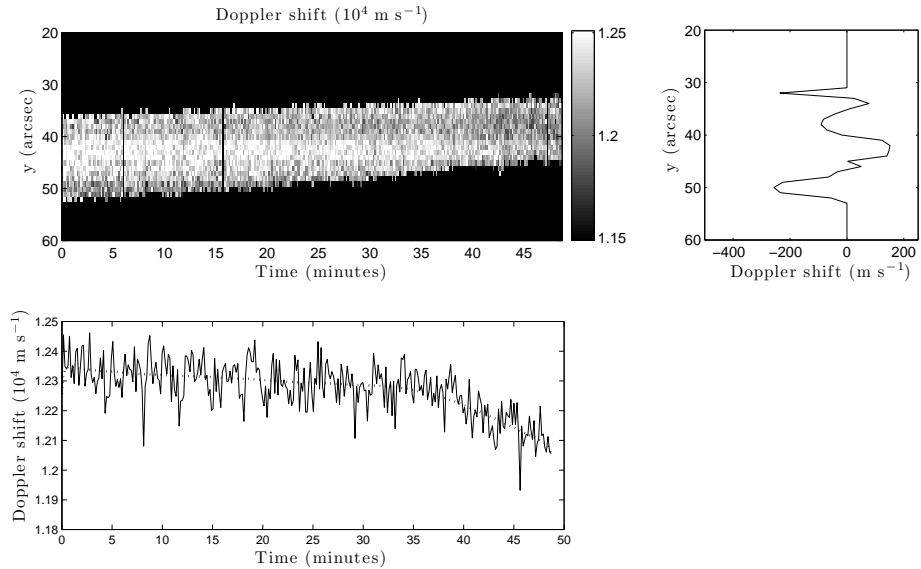}
\caption{Top left: Doppler shift diagram as a function of time
($x$-axis) and space ($y$-axis). Doppler shift is expressed in $10^4$  m
s$^{-1}$. Bottom left: mean Doppler shift with time. Top right:
mean Doppler shift with spatial dimension. The dashed line on the
bottom left plot indicates the fitted estimate obtained by a
weighted moving average, which has been used to characterize the
vertical distorsion of the Doppler ``surface'' plotted in bottom
right figure.} \label{fig:les_0}
\end{figure}

\begin{figure}
\center
\includegraphics[width=11cm]{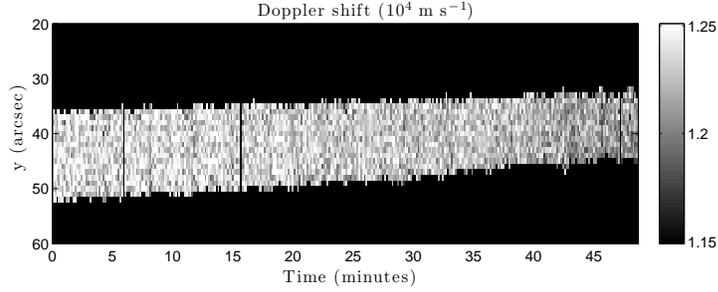}
\caption{Clean Doppler diagram, obtained after subtraction of the
spurious distorted ``surface'' enlightened in the raw diagram (Fig
\ref{fig:les_0}).} \label{fig:les_0_propres}
\end{figure}

The global measurement of the Doppler shift is represented in a
spatial-temporal diagram on Fig. \ref{fig:les_0}, top left.
Although the scan motion was perpendicular to terminator, we note
that the upper edge of the planet is a straight line, whereas the
bottom edge is curved. This is due to the fact that manual guiding
was performed on the top of Venus image in the field (southern
hemisphere). This guiding procedure and resulting vertical drift
allowed us to reveal a spectral artifact probably due to the
missing calibration procedures we referred to in section 2, \S
2.3. This spectral artifact is visible on Fig. \ref{fig:les_0},
top left frame, as a white horizontal band between y = 40 and y =
45 arcsec. We decided to pursue the analysis by trying to correct
it the following way. First, we coadded all the lines to assess
the mean variation of the Doppler integrated over slit height.
This mean variation is shown in  Fig. \ref{fig:les_0}, bottom. It
shows qualitatively that the spectral shift between Earth's solar
D2 and Venus' solar D2 decreases as the slit moves away from
terminator. This average decrease is then fitted to a weighted
moving average in order to flatten the Doppler surface, with
respect to time, prior to the kinematical fit described in Section
4. Second, the temporal mean of this diagram, shown in  Fig.
\ref{fig:les_0}, right frame, has been subtracted to the main data
frame in order to remove the artifact. The result is shown in Fig.
\ref{fig:les_0_propres}.

\section{Results}
\label{results}
\subsection{Working with relative velocities}
Strong discrepancies of up to several tens of m s$^{-1}$ have been
commonly met with tentatives of making absolute radial velocity
measures using visible lines. This has been discussed in the case
of Venus wind measurements in Young et al. (1979), Widemann et al.
(2007, 2008). These authors concluded for the need of a reference
point on Venus used as a relative velocity reference, and they
used this point to perform differential velocity measurements on
the disk. The 0 velocity is fixed at the planetary coordinates
$(\theta = 0^\circ, \alpha = 5^\circ)$, where $\theta$ and
$\alpha$ indicate the latitude and longitude.

\begin{figure}
\center
\includegraphics[width=6.5cm]{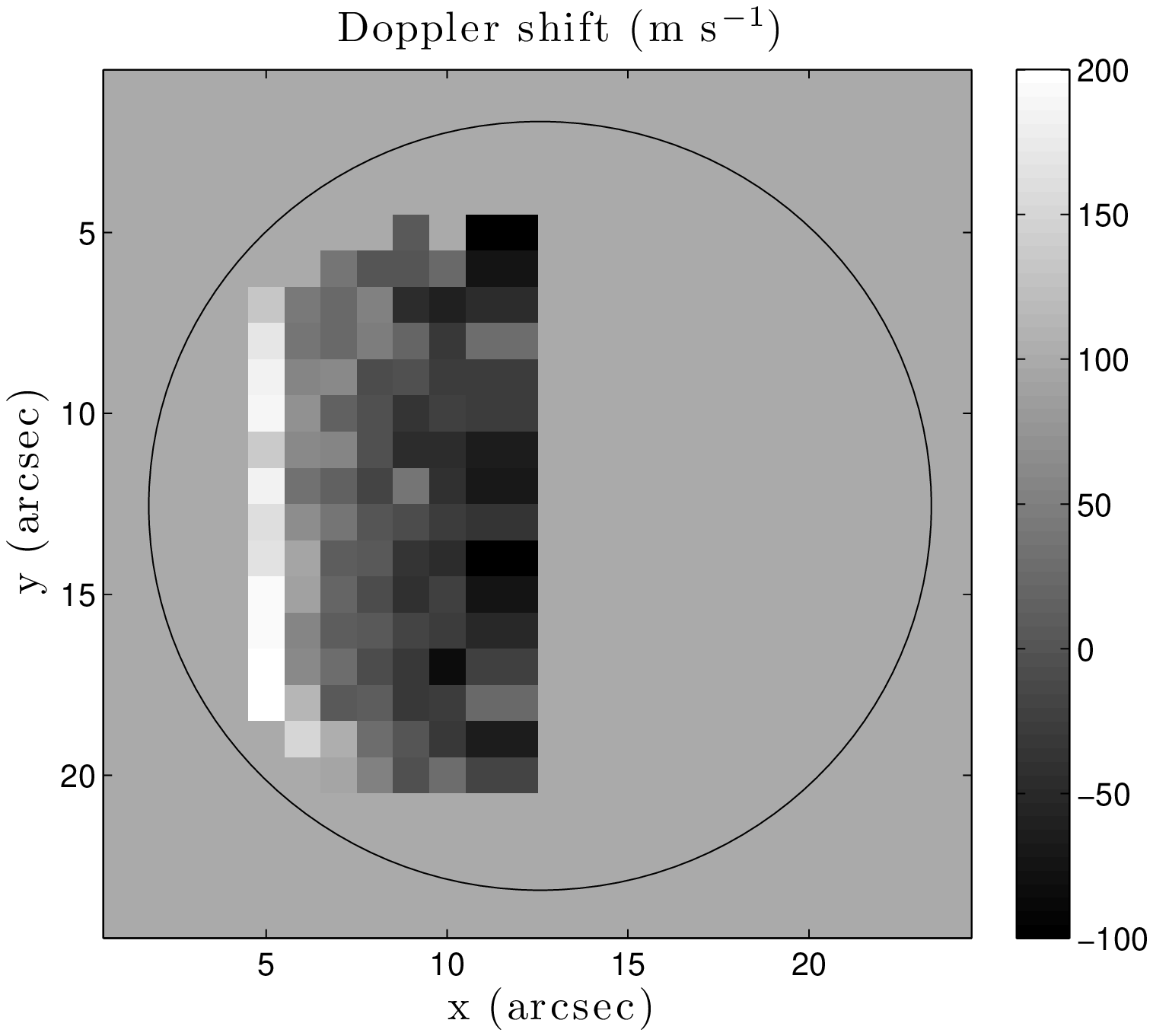}
\includegraphics[width=6.5cm]{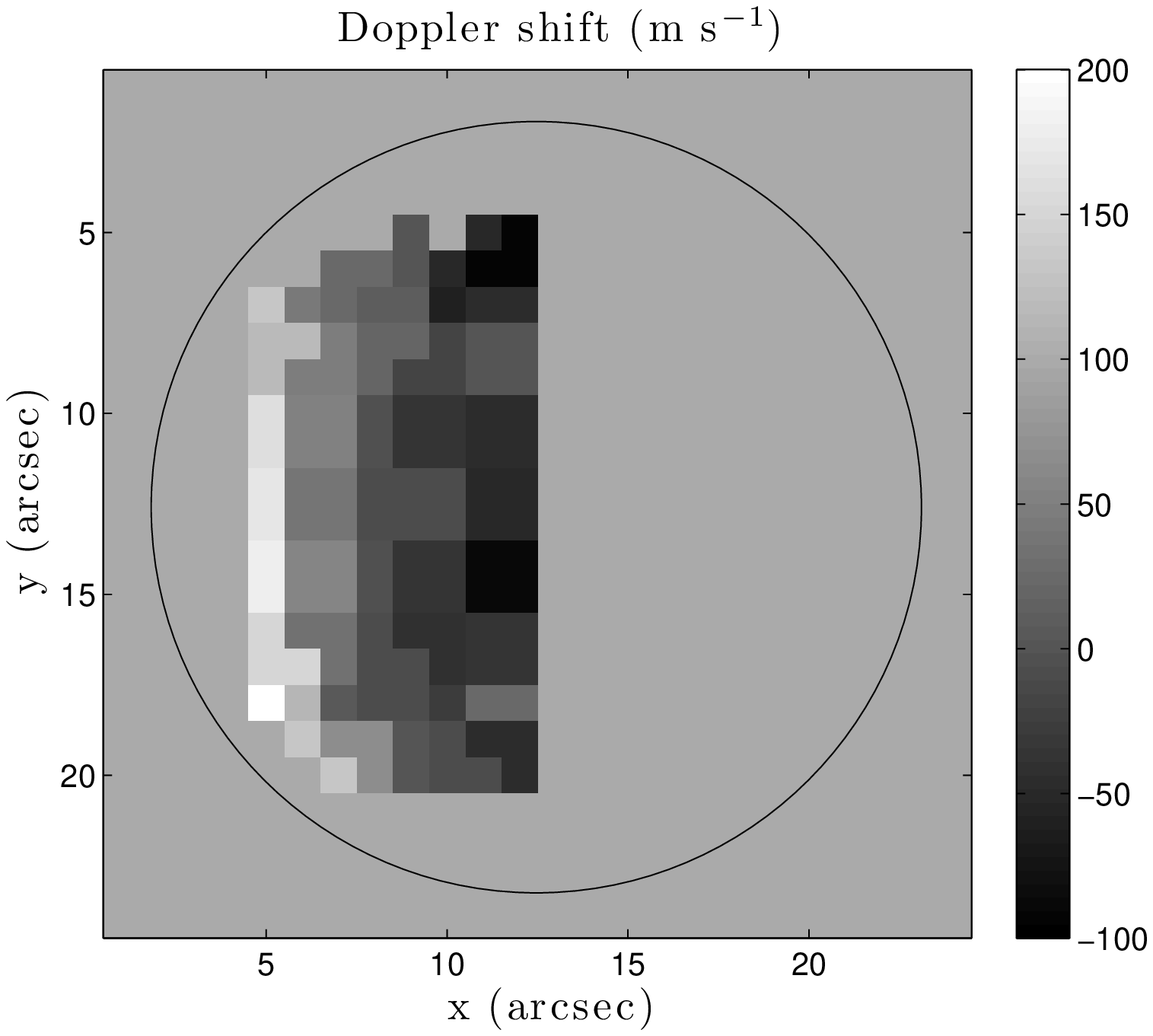}
\caption{Left: Relative velocity map obtained after summation of
Doppler shift within 1-arcsec intervals. The dot line circle
indicates the planetary diameter. The maximum latitude extension
reaches $\pm45^\circ$, while the maximum longitude reaches 55° at
pixels $(5,7)$ and $(5,17)$. Right: the same Doppler map where
pixels have been averaged within a regular latitude-longitude
grid, spaced by 10$^\circ$. Latitude range is
$[-45^\circ,45^\circ]$ while longitude range is
$[0^\circ,55^\circ]$.} \label{fig:doppler_map}
\end{figure}

The radial velocity map is shown in Fig. \ref{fig:doppler_map}.
The maximum velocity difference on the whole map reaches almost
300 m s$^{-1}$, while the mean amplitude of the variation of
velocity across the planet is about 200 m s$^{-1}$. A rough
estimate of the actual mean noise level can be obtained by measuring the
standard deviation $\Sigma$ of each ``column of pixels'' on Venus figure. The
mean value of the dispersion of points along a column is equal to 31 m s$^{-1}$ (see Tab
\ref{tab:std_doppler_map}). Note that the higher noise level, which is observed
in column of abscissa $x=[5,6]$, is due to a geometrical effect.
Indeed, the wind
velocity on Venus varies more strongly in columns far from the
terminator, because a wide range of longitude is explored, what increases the standard deviation of the considerer column. This fact makes the mean standard deviation value appear as a slightly pessimistic estimate of the actual mean noise level per pixel. Nevertheless, its value, about 31 m s$^{-1}$, almost squares with the theoretical performance (30.7 m s$^{-1}$) presented in Sect. \ref{theo_perf}, what shows that our estimate of the noise level is correct.

\subsection{Fit of Doppler winds to zonal circulation}
Doppler blueshift between Venus' atmosphere and the Sun is maximum along the
terminator, whereas Doppler redshift between Earth and Venus is
maximum along the planetary limb. By supposing a purely zonal
wind, the isotach corresponding to radial velocity $v\ind{rad} =
0$ is the meridian defined by the bisecting angle between
sub-earth and sub-solar longitudes. Moreover, a correction
has been introduced by Young et al. (1979), taking into account
the solar apparent diameter and its rotation
seen from Venus (42 arcmin), so-called {\it Young effect} (see also Widemann et al., this issue). 

The consequence is
an increase of the apparent Doppler shift for the observer, along the terminator
at mid and high latitudes ($\pm 45^\circ$). The typical
amplitude of the wind increase reaches almost 30 m
s$^{-1}$ for a 100 m s$^{-1}$ zonal wind, and therefore must be arbitrarily corrected in
a kinematical fit to a pure zonal regime.

The algorithm used to extract the velocity amplitude from the
radial velocity map has been adapted from Widemann et al. (2007).
The zonal circulation at cloud top level is characterized by a latitudinal
dependency and wind decrease in the polar regions. Recently reanalysed
Pioneer Venus UV data (Limaye, 2007) and SSI Galileo imaging (Peralta et
al. 2007) indicate a generally  uniform velocity between latitude $\pm 50^\circ$
with a best fit to a constant angular velocity at higher latitudes, in
accordance with winds measured from cloud tracking by both VIRTIS-M and
VMC (Markiewicz et al., 2007 ; Piccioni et al., 2007). Both types of zonal
wind regimes dependency have been applied to fit our data using classical
least-square algorithm.

For a uniform, solid body circulation, the wind velocity is estimated at
at 2-$\sigma$ at $151 \pm 16$ m s$^{-1}$, with reduced $\chi^2 =
1.69$. On the other hand, with a cosine latitudinal dependance,
the zonal wind velocity is estimated at $146 \pm 17$ m s$^{-1}$, with
reduced $\chi^2 = 1.85$. The close results between the two
approaches is due to the fact that our observations do not explore
(with good SNR) Venus wind map at latitude higher than $45^\circ$.
Under this latitude, the difference between both models is not very
significant. It has to be noticed that uncertainties on the wind global velocity (16-17 m s$^{-1}$) is larger that what would be expected from local noise level. Indeed,  with a local noise level of 31 m s$^{-1}$ per 1 arcsec square pixel on about 128 pixels, the global noise level by integrating all Venus' pixels would decrease to $31/\sqrt{128} = 2.7$ m s$^{-1}$. This lower performance is due to three facts. First, the sensitivity of Doppler measurements to velocity is not uniform on the planet (isotachs are functions of longitude, e. g. Widemann et al. 2007). Second, the real velocity field might not be uniform as supposed inside the model used to fit the data. At last, the previously exposed uncertainty about the positioning introduces a bias in the global fit. 

Our result is compatible with Doppler spectroscopy measurements of Widemann et al. (2007), where the
wind amplitude is estimated in a [90, 150] m s$^{-1}$ velocity range. However, this result represents an upper value with respect to VEx results of Markiewicz et al. (2007), who have obtained a mean value zonal wind of 95 m/s between latitudes 10¼N and 40¼S, using cloud tracking method. Two reasons may explain the discrepancy between our results and those of Markiewicz et al. (2007). First, it could be a consequence of the uncertainty on the positioning, what would implies a 60-m s$^{-1}$ bias in global velocity estimate. Second, it might point out the fact that cloud tracking and Doppler spectrometry are two distinct approaches to measure the wind velocity. With cloud tracking, one measure the cloud feature motion, while Doppler spectrometry measures the cloud particle motion, which may differ.  

\section{Conclusions and prospects}
The goal of our observations was to evaluate the ability of the
MTR/THEMIS solar telescope in order to measure velocity wind by
Doppler spectroscopy in the visible range. Despite cloudy weather,
and consequently a very short run (49 min), we obtain a promising
instrumental performance:  the mean noise level on the velocity map is about 31 m
s$^{-1}$ per 1-arcsec pixel, which corresponds to the expectations. As regards the wind velocity field, it has been estimated at 149 $\pm$ 16 m s$^{-1}$, what represents a quiet excessive value with respect to other observations. We have given two possible explanations. First, it could come from a global bias introduced by the observation conditions, in particular to the lack of quick
scan, which has made the positioning upon the planetary disk
noisy. Also, It might be due to the fact that cloud tracking and Doppler measurements represent different approaches to wind velocity estimate. Part of this spurious velocity which has be seen in fig. 10
would be skipped out with the use of the tip-tilt guiding system  
and by optimizing the instrumental configuration. By supposing the same local noise level (31 m s$^{-1}$) and by choosing Fraunhoffer lines with a better sensitivity to Doppler shifts, it would be possible to reach a noise level around 10 m s$^{-1}$ per 1 arcsec square in few hour observation run and to reach a global wind measurement accuracy of about several m s$^{-1}$.

These encouraging performance have motivated a new observation
campaign, which is planned for mid spring 2008. Tip tilt guiding
will be used. Venus will present a shorter apparent diameter (10
arcsec) and a greater phase (90 \%). Four deep solar lines will be
used to measure velocity fields (Fe I, Mg, D1 and D2 Na), in order
to gain a factor 2 in the SNR. Moreover, the Doppler shift of
CO$_2$ line ($\nu_3$ band, 8680 \AA) will be studied in order to
probe 7 km higher. Future observations will be of major
interest because at short elongation Venus is practically
unobservable with classical night telescopes, whereas VEx is still
orbiting around Venus.




\newpage

\begin{table}[h!]
\centering
 \caption{Observation properties of Venus on November 7th 2008
 from Teide Observatory (IMCCE ephemeris database).}
\begin{tabular}{|l|c|c|c|c|c|c| }\hline
Date UTC  & R.A & Dec.& Distance & V.Mag & Phase  & Dist dot\\
           h  m s    &   h  m  s       &  o  '  "    &  ua. & & o &  o \\
           \hline
13  6  0.00  & 11 56 21.86 & 01 28  42.10&
0.77 & -4.31  & 83.58  & 13.27725\\
13 56 0.00 &  11 56 30.16& 01 27 57.89 &0.77&  -4.31&
83.56& 13.32962\\
\hline
\end{tabular}
\label{carlossanchez}
\end{table}

\vspace{1cm}

\begin{table}[h!]
\centering
 \caption{Standard deviation $\Sigma$ of velocity in each column of the Venus map.
 The mean standard deviation value is equal to 31 m s$^{-1}$.
 The third line indicate the number of spectra averaged in order to build each column.
 We have averaged column 11 and 12 because column 11 alone have only 5 spectra,
 which affect the global noise level. The standard deviation increases in column 5 and 6
 and 7 despite a major number of averaged spectra, because of the wind velocity strong
 variation in these columns. }
\begin{tabular}{|l|c|c|c|c|c|c|c| }\hline
Abcissa $x$ & (12,11) & 10 & 9 & 8& 7 & 6&5\\
\hline $\Sigma$ (m s$^{-1}$) & 42.96 & 27.60 & 25.06 & 23.93 &
30.27& 33.00 & 34.72\\\hline
Number of spectra & 30 & 31 & 40 & 48 & 56 & 67 & 43\\
 \hline
\end{tabular}
\label{tab:std_doppler_map}
\end{table}

\newpage


\end{document}